# COMMENTS ON HIGH-DIMENSIONAL SIMULTANEOUS INFERENCE WITH THE BOOTSTRAP

## JELENA BRADIC AND YINCHU ZHU

The authors should be congratulated on their insightful article proposing forms of residual and paired bootstrap methodologies in the context of simultaneous testing in sparse and high-dimensional linear models. We appreciate the clear exposition of their work, and the effectiveness of the proposed method. The authors advocate for the bootstrap of a complete high-dimensional estimate rather than the linearized part of the test statistic. We appreciate the opportunity to comment on several aspects of this article.

## 1. BOOTSTRAPS RELATIVE EFFICIENCY

The problem of forming a confidence set for the many parameters of interests in high-dimensional setting differs from a more routine interval estimation problems in low-dimensional setting, in that the estimator itself induces shrinkage and bias at estimation. These difficulties, in turn, prohibit simple and intuitive guidelines for judging the relative efficiency of one bootstrap scheme with respect to the other. Thus it seems interesting to discuss relative efficiencies measured by the width of the confidence intervals (among those that achieve nominal coverage).

We performed a small scale simulation study to investigate a number of scenarios. For that end, we consider $Y = X\beta^* + \varepsilon$, where $X \in \mathbb{R}^{n \times p}$ has i.i.d entries generated from $\mathcal{N}(0,1)$ and $\varepsilon \in \mathbb{R}^n$ has i.i.d entries across $i$ with $\varepsilon_i = \sigma_{\varepsilon,i} e_i$. Here $e_i$ is independent of $\sigma_{\varepsilon,i}$ and follows one of the four distributions: (1) $\mathcal{N}(0,1)$ (Gaussian), (2) centered exponential random variables with parameter 1 (Exponential), (3) student-t distribution with 6 degrees of freedom normalized to have variance equal to 1 (Student) and (4) mixture of two Gaussian distributions centered at 0.95 and -0.99 such that the mixture distribution has mean zero and variance one (Mixture). The conditional standard deviation $\sigma_{\varepsilon,i}$ is either 1 (homoscedastic) or $|X_{i,2}|$ (heteroscedastic). We use $\beta^* = (1, 1, 0, ..., 0)^\top \in \mathbb{R}^p$. The goal is to test

$$H_0: \quad \beta_j^* = \beta_j^o \qquad j = \{2, ..., 80\}.$$

We report coverage probability and width of 95% confidence sets. For the ease of comparison across different sample size, the width times $\sqrt{n}$ is reported. The results are computed using 2000 random samples. We write Residual bootstrap (RB), multiplier wild bootstrap with Gaussian multipliers (MBG), xyz-paired bootstrap (xyz), Zhang and Cheng (ZC), robust Zhang and Cheng (RZC), multiplier bootstrap with Radmacher multipliers (MBR) and multiplier bootstrap with Mammen multipliers (MBM). For robust Zhang and Cheng, instead of bootstrapping $\|n^{-1/2} \sum_{i=1}^n \hat{\Theta}^\top X_i\|_\infty$ and multiplying it by $\hat{\sigma}_\varepsilon$ as proposed by [3], we directly bootstrap $\|n^{-1/2} \sum_{i=1}^n \hat{\Theta}^\top X_i \hat{\varepsilon}_i\|_\infty$, where $\hat{\Theta}$ is the nodewise Lasso estimator as defined therein.

Table 1. The Coverage (Cov) and Width of the confidence interval. Design has Gaussian distribution whereas the distribution of the errors is homoscedastic and varies from symmetric to non-symmetric to heavy tailed to bimodal.

| | | | | | | | | |
|---|---|---|---|---|---|---|---|---|
| | | | | $n = 100$ and $p = 150$ | | | | |
| | Gaussian | | Exponential | | Student | | Mixture | |
| | Cov | Width | Cov | Width | Cov | Width | Cov | Width |
| RB | 0.947 | 3.382 | 0.952 | 3.337 | 0.944 | 3.336 | 0.936 | 3.418 |
| MBG | 0.922 | 3.269 | 0.934 | 3.237 | 0.925 | 3.235 | 0.904 | 3.316 |
| MBR | 0.943 | 3.373 | 0.950 | 3.323 | 0.941 | 3.322 | 0.934 | 3.407 |
| MBM | 0.909 | 3.253 | 0.925 | 3.221 | 0.913 | 3.222 | 0.901 | 3.298 |
| XYZ | 0.990 | 3.809 | 0.991 | 3.834 | 0.990 | 3.840 | 0.976 | 3.805 |
| ZC | 0.945 | 3.838 | 0.955 | 3.814 | 0.948 | 3.775 | 0.942 | 3.844 |
| RZC | 0.967 | 3.959 | 0.974 | 4.053 | 0.972 | 4.018 | 0.952 | 3.907 |
| | | | | $n = 200$ and $p = 300$ | | | | |
| RB | 0.943 | 3.366 | 0.947 | 3.322 | 0.939 | 3.327 | 0.941 | 3.403 |
| MBG | 0.924 | 3.285 | 0.928 | 3.247 | 0.931 | 3.246 | 0.925 | 3.335 |
| MBR | 0.939 | 3.366 | 0.944 | 3.314 | 0.950 | 3.321 | 0.936 | 3.400 |
| MBM | 0.921 | 3.291 | 0.934 | 3.248 | 0.927 | 3.250 | 0.924 | 3.333 |
| XYZ | 0.967 | 3.591 | 0.985 | 3.626 | 0.979 | 3.635 | 0.966 | 3.576 |
| ZC | 0.940 | 3.624 | 0.951 | 3.595 | 0.946 | 3.595 | 0.948 | 3.626 |
| RZC | 0.953 | 3.689 | 0.966 | 3.760 | 0.963 | 3.766 | 0.960 | 3.646 |

1.1. **Gaussian design.** What we have observed in this simulation study differs from the classical low-dimensional setting where wild bootstrap and paired bootstrap provide the most robust bootstrap alternatives. Table 1 contains results of the simulation study and indicate that the MBG and MBM undercover and that XYZ bootstrap has much larger coverage than the nominal level of 95%. The effects of shrinkage, the nonlinearity it induces and the tuning parameter choices, result in pattern of the bootstrap efficiency that is different from the intuitive one (see [1], Section 2.1). The MBR and RB perform similarly with coverage of around 95% and the shortest widths among all the methods. In low-dimensional settings, [2] shows that the distribution of the wild bootstrap converges faster than the paired bootstrap; however, we did not observe any indication that this may be true in high-dimensions. The proposed ZC and RZC even in the case of nominal coverage provide much wider confidence intervals suggesting they are less efficient in this example. With the increase in the sample size, the difference in the width shrunk suggesting possible slower convergence rate or RZC and ZC in comparison to MBR and RB. On the other hand, increase in the sample size improved the coverage of XYZ; however, it did not improve the coverage of MBG and MBM. Lastly, all seven methods were robust to the size of the $p$.

1.2. **Heteroscedasticity.** The wild bootstrap has often been interpreted as a procedure that resamples residuals in a manner that captures any heteroscedasticity in the underlying errors. Section 5.2 in [1] gives Monte Carlo evidence supporting the superiority of the wild bootstrap for carrying out a t test for the least squares estimator in the heteroskedastic linear model. A central thesis of the article is that the failure of existing multiplier bootstrap schemes, such as the multiplier bootstrap [3], is due to its neglect of the excess variation resulting from the possible non-gaussianity of the model error or the presence of heteroscedastic errors.

Table 2. The Coverage (Cov) of the confidence interval and width at two significances: W90 for 90% and W95 for 95%. Design has Gaussian distribution whereas the distribution of the errors is heteroscedastic and varies from symmetric to non-symmetric to heavy tailed to lastly bimodal.

| | Gaussian | | Exponential | | Student | | Mixture | |
|---|---|---|---|---|---|---|---|---|
| | Cov | Width | Cov | Width | Cov | Width | Cov | Width |
| | | | $n = 100$ and $p = 150$ | | | | | |
| RB | 0.946 | 3.306 | 0.939 | 3.240 | 0.960 | 3.284 | 0.943 | 3.341 |
| MBG | 0.923 | 3.199 | 0.920 | 3.155 | 0.939 | 3.184 | 0.924 | 3.234 |
| MBR | 0.942 | 3.278 | 0.930 | 3.195 | 0.954 | 3.250 | 0.943 | 3.322 |
| MBM | 0.915 | 3.187 | 0.914 | 3.152 | 0.931 | 3.160 | 0.919 | 3.232 |
| XYZ | 0.994 | 3.883 | 0.989 | 3.958 | 0.994 | 3.925 | 0.989 | 3.920 |
| ZC | 0.910 | 3.788 | 0.914 | 3.735 | 0.927 | 3.679 | 0.923 | 3.803 |
| RZC | 0.957 | 4.216 | 0.960 | 4.351 | 0.970 | 4.160 | 0.956 | 4.164 |
| | | | $n = 200$ and $p = 300$ | | | | | |
| RB | 0.942 | 3.304 | 0.946 | 3.233 | 0.950 | 3.268 | 0.943 | 3.346 |
| MBG | 0.919 | 3.213 | 0.923 | 3.155 | 0.927 | 3.181 | 0.917 | 3.254 |
| MBR | 0.934 | 3.291 | 0.943 | 3.216 | 0.950 | 3.251 | 0.938 | 3.343 |
| MBM | 0.917 | 3.218 | 0.924 | 3.170 | 0.928 | 3.183 | 0.921 | 3.269 |
| XYZ | 0.982 | 3.640 | 0.985 | 3.714 | 0.989 | 3.709 | 0.977 | 3.615 |
| ZC | 0.912 | 3.592 | 0.898 | 3.550 | 0.920 | 3.562 | 0.904 | 3.600 |
| RZC | 0.960 | 3.966 | 0.956 | 4.083 | 0.969 | 4.048 | 0.940 | 3.904 |

As claimed by the authors, ZC method undercovers in all heteroscedastic cases. Our proposed robust version of the ZC, RZC, put the coverage at the correct order; however, it creates confidence intervals which are prominently wide. The XYZ method over covers but interestingly has the width much smaller than the RZC, indicating suboptimality of the RZC method. However, traditional benefits of the wild bootstrap appear lost now as the wild bootstrap is not able to capture the heteroscedasticity; MBG and MBM both under cover significantly below the expected level.

1.3. **Non-Gaussian design.** In this section we considered a non-gaussian design model and tested the ability of the proposed methods to adapt. We observed that both RB and ZC showcase strong robustness to the distribution of the design; both methods relate to each other in the similar way to Section 1.1. However, here we observed the MBR is not always stable with the non-gaussian design with MBG and MBM failing to cover at the nominal level. XYZ is still over covering.

Table 3. The Coverage (Cov) and Width of the confidence interval. Design has Exponential distribution whereas the distribution of the errors is homoscedastic and varies from symmetric to non-symmetric to heavy tailed to bimodal.

$n = 100$ and $p = 150$

|  | Gaussian | | Exponential | | Student | | Mixture | |
|---|---|---|---|---|---|---|---|---|
|  | Cov | Width | Cov | Width | Cov | Width | Cov | Width |
| RB | 0.933 | 3.309 | 0.951 | 3.325 | 0.946 | 3.279 | 0.943 | 3.338 |
| MBG | 0.914 | 3.223 | 0.916 | 3.206 | 0.933 | 3.199 | 0.931 | 3.268 |
| MBR | 0.931 | 3.303 | 0.930 | 3.271 | 0.947 | 3.269 | 0.947 | 3.350 |
| MBM | 0.910 | 3.202 | 0.908 | 3.196 | 0.925 | 3.180 | 0.922 | 3.251 |
| XYZ | 0.986 | 3.898 | 0.993 | 3.940 | 0.993 | 3.939 | 0.988 | 3.887 |
| ZC | 0.946 | 4.091 | 0.932 | 4.082 | 0.934 | 4.047 | 0.957 | 4.121 |
| RZC | 0.972 | 4.387 | 0.984 | 4.667 | 0.982 | 4.633 | 0.971 | 4.276 |

$n = 200$ and $p = 300$

| RB | 0.941 | 3.301 | 0.942 | 3.368 | 0.939 | 3.275 | 0.940 | 3.335 |
|---|---|---|---|---|---|---|---|---|
| MBG | 0.916 | 3.233 | 0.893 | 3.204 | 0.918 | 3.195 | 0.925 | 3.279 |
| MBR | 0.934 | 3.299 | 0.906 | 3.265 | 0.932 | 3.258 | 0.938 | 3.346 |
| MBM | 0.925 | 3.235 | 0.880 | 3.206 | 0.922 | 3.202 | 0.922 | 3.274 |
| XYZ | 0.983 | 3.653 | 0.980 | 3.746 | 0.985 | 3.713 | 0.975 | 3.614 |
| ZC | 0.949 | 3.765 | 0.920 | 3.738 | 0.930 | 3.747 | 0.947 | 3.769 |
| RZC | 0.968 | 3.964 | 0.982 | 4.252 | 0.976 | 4.280 | 0.956 | 3.846 |

1.4. **Non-gaussian design and heteroscedasticity.** The effect of the heteroscedasticity here is more pronounced as the design is non-gaussian. We observe the complete failure of the ZC method with coverage going as low as 70% for the nominal level of 95%. In this case, we observe that RB needs larger sample size to cover the heavy-tailed error distributions; however, most methods underperform substantially. All of the wild multiplier bootstraps fail with MBR slightly performing better for larger $n$. RZC perhaps has the most consistency in covering, but its width can be massive. The case of the bimodal error seems to be particularly difficult for all of the methods and none of them perform sufficiently well. Interestingly, we have not found a case where XYZ performs much better than the RB or MBR.

## 2. Bootstrap inference for high-dimensional and possibly non-sparse models

The interesting and shared part of many of the proposed bootstrap schemes is in the construction of the residuals based on regularized, i.e., Lasso estimator. The most prominent examples, exhibiting excellent performance in a variety of settings, seem robust to the construction of such residuals despite the intricate bias introduced by the regularization. However, when the sparsity of the model increases, we expect that the introduced bootstrap, that is the Lasso within it, will induce a bias term that would be too large and that would affect the finite sample coverage.

TABLE 4. The Coverage (Cov) and Width of the confidence interval. Design has Exponential distribution whereas the distribution of the errors is heteroscedastic and varies from symmetric to non-symmetric to heavy tailed to bimodal.

| | Gaussian | | Exponential | | Student | | Mixture | |
|---|---|---|---|---|---|---|---|---|
| | Cov | Width | Cov | Width | Cov | Width | Cov | Width |
| | | | $n = 100$ and $p = 150$ | | | | | |
| RB | 0.944 | 3.249 | 0.902 | 3.207 | 0.939 | 3.237 | 0.913 | 3.270 |
| MBG | 0.922 | 3.170 | 0.882 | 3.137 | 0.917 | 3.150 | 0.891 | 3.207 |
| MBR | 0.941 | 3.275 | 0.896 | 3.199 | 0.935 | 3.239 | 0.914 | 3.358 |
| MBM | 0.925 | 3.176 | 0.885 | 3.169 | 0.919 | 3.159 | 0.902 | 3.249 |
| XYZ | 0.995 | 4.149 | 0.964 | 4.308 | 0.993 | 4.181 | 0.972 | 4.404 |
| ZC | 0.797 | 3.871 | 0.756 | 3.757 | 0.802 | 3.775 | 0.749 | 3.911 |
| RZC | 0.950 | 5.154 | 0.933 | 5.283 | 0.955 | 5.127 | 0.906 | 5.108 |
| | | | $n = 200$ and $p = 300$ | | | | | |
| RB | 0.943 | 3.242 | 0.921 | 3.202 | 0.951 | 3.223 | 0.927 | 3.274 |
| MBG | 0.934 | 3.164 | 0.909 | 3.127 | 0.933 | 3.146 | 0.913 | 3.206 |
| MBR | 0.943 | 3.238 | 0.921 | 3.180 | 0.953 | 3.211 | 0.931 | 3.306 |
| MBM | 0.929 | 3.175 | 0.910 | 3.151 | 0.929 | 3.153 | 0.919 | 3.241 |
| XYZ | 0.993 | 3.887 | 0.978 | 4.056 | 0.993 | 3.942 | 0.984 | 4.001 |
| ZC | 0.737 | 3.641 | 0.705 | 3.584 | 0.737 | 3.555 | 0.727 | 3.670 |
| RZC | 0.926 | 5.194 | 0.934 | 5.346 | 0.944 | 5.165 | 0.901 | 5.198 |

We explore the possibility of applying the bootstrap methods proposed by the authors to perform inference problems of high-dimensional models that are potentially non-sparse. To the best of our knowledge, the only work in this direction is [4]. We only consider the particular case of $G$ being a singleton, i.e., testing one entry of $\beta$. Without loss of generality, consider the problem of testing

$$H_0 : \ \beta_1^* = \beta_1^o.$$

The CorrT method proposed in [4] is implemented as follows. We first compute

$$\hat{\beta}_{-1} = \arg\min_{v \in \mathbb{R}^{p-1}} \|v\|_1$$
$$s.t. \quad \|X_{-1}^\top (Y - X_1 \beta_1^o - X_{-1} v)\|_\infty \le \eta_\beta$$
$$\|Y - X_1 \beta_1^o - X_{-1} v\|_\infty \le \|Y - X_1 \beta_1^o - X_{-1} v\|_2 / \log^2 n$$
$$(Y - X_1 \beta_1^o)^\top (Y - X_1 \beta_1^o - X_{-1} v) \ge 0.01 \|Y - X_1 \beta_1^o\|_2^2 / \sqrt{\log n},$$

and

$$\begin{align}
\hat{\theta} &= \arg\min_{v \in \mathbb{R}^{p-1}} \|v\|_1 \tag{1}\\
s.t. &\quad \|X_{-1}^\top (X_1 - X_{-1} v)\|_\infty \le \eta_\theta\\
&\quad \|X_1 - X_{-1} v\|_\infty \le \|X_1 - X_{-1} v\|_2 / \log^2 n\\
&\quad X_1^\top (X_1 - X_{-1} v) \ge 0.01 \|X_1\|_2^2 / \sqrt{\log n},
\end{align}$$

where $\eta_\beta, \eta_\theta \asymp \sqrt{n^{-1} \log p}$ are tuning parameters chosen as in [4]. Let

$$\hat{u} = X_1 - X_{-1}\hat{\theta}, \text{ and } \hat{\varepsilon} = Y - X_1\beta_1^o - X_{-1}\hat{\beta}_{-1}.$$

Then the test statistic defined therein takes the form

$$T_n = \sqrt{n} \ \hat{\varepsilon}^\top \hat{u} / \|\hat{\varepsilon}\|_2 \|\hat{u}\|_2.$$

Under the assumptions in [4], it is proved that under $H_0 : \beta_1^* = \beta_1^o$, $T_n$ converges in distribution to $\mathcal{N}(0, 1)$, regardless of whether or not $\beta^*$ is sparse. In the following we define a residual bootstrap counterparts of the test statistic above and explore their finite sample properties.

### 2.1. **CorrT with residual bootstrap (CorrTRB).**
Define centered estimated residuals with $\bar{u} = (\bar{u}_1, ..., \bar{u}_n)^\top$ with $\bar{u}_k = \hat{u}_k - n^{-1} \sum_{i=1}^n \hat{u}_i$. Then, consider a random sample $u^{RB} = (u_1^{RB}, ..., u_n^{RB})^\top$ computed by drawing with replacement from $\{\bar{u}_i\}_{i=1}^n$. Then, define the new variable $X_1^{RB} = X_{-1}\hat{\theta} + u^{RB}$. We then compute $\hat{\theta}^{RB}$ as $\hat{\theta}$ in (1) with $X_1$ replaced by $X_1^{RB}$. The bootstrap test statistic is

$$T_n^{RB} = \frac{\sqrt{n}\hat{\varepsilon}^\top (X_1^{RB} - X_{-1}\hat{\theta}^{RB})}{\|\hat{\varepsilon}\|_2 \|X_1^{RB} - X_{-1}\hat{\theta}^{RB}\|_2}.$$

### 2.2. **CorrT with paired bootstrap (CorrTPB).**
For this setting we would like to take advantage of the distributions of pairs of observations. We begin by defining

$$\hat{X}_1 = \hat{X}_{-1}\hat{\theta} + \bar{u}$$

and $\hat{X}_{-1} = (\hat{X}_2, ..., \hat{X}_p) \in \mathbb{R}^{n \times (p-1)}$ with

$$\hat{X}_j = X_j - \|\bar{u}\|_2^{-2}(X_j^\top \bar{u})\bar{u}$$

for $j \geq 2$. We now consider a new sample $(\hat{\varepsilon}^*, \hat{X}_1^*, \hat{X}_{-1}^*)$ whose rows form a random sample with replacement from rows of $(\hat{\varepsilon}, \hat{X}_1, \hat{X}_{-1})$. We compute $\hat{\theta}^*$ as $\hat{\theta}$ in (1) with $(X_1, X_{-1})$ replaced by $(\hat{X}_1^*, \hat{X}_{-1}^*)$. The bootstrap test statistic reads

$$T_n^{pair} = \frac{\sqrt{n}\hat{\varepsilon}^{*\top}(X_1^* - X_{-1}^*\hat{\theta}^*)}{\|\hat{\varepsilon}^*\|_2 \|X_1^* - X_{-1}^*\hat{\theta}^*\|_2}.$$

### 2.3. **CorrT with wild multiplier bootstrap (CorrTMB).**
Let $\{\xi_i\}_{i=1}^n$ be i.i.d. multipliers drawn independent of the sample. We compute $X_1^{MB} = X_{-1}\hat{\theta} + u^{MB}$, where $u^{MB} = (u_1^{MB}, ..., u_n^{MB})^\top$ with

$$u_i^{MB} = \bar{u}_i \xi_i.$$

Then we compute $\hat{\theta}^{MB}$ as $\hat{\theta}$ in (1) with $X_1$ replaced by $X_1^{MB}$ and the bootstrap test statistic

$$T_n^{MB} = \frac{\sqrt{n}\hat{\varepsilon}^\top (X_1^{MB} - X_{-1}\hat{\theta}^{MB})}{\|\hat{\varepsilon}\|_2 \|X_1^{MB} - X_{-1}\hat{\theta}^{MB}\|_2}.$$

We propose three multiplier bootstrap methods labeled by CorrTMBG, CorrTMBR and CorrTMBM, where the multipliers are drawn from $\mathcal{N}(0, 1)$, Radmacher distribution and Mammen distribution, respectively.

2.4. **Numerical example.** We now compare CorrT and its bootstrap variations. We report the rejection probabilities of these methods for the hypothesis $H_0 : \beta_1^* = \beta_1^o$ with $\beta_1^o = \beta_1^* + n^{-1/2}h$. Hence, these probabilities represent the size if $h = 0$ and the power if $h \neq 0$.

TABLE 5. CorrT: Gaussian design and Gaussian errors. We consider dense $\beta^* = 4p^{-1/2}(1, ..., 1)^\top \in \mathbb{R}^p$ with $p = 300$ and $n = 200$.

| | | | Homoscedasticity | | | |
| | CvM | KS | $h = 0$ | $h = 5$ | $h = 10$ | $h = 15$ |
|---|---|---|---|---|---|---|
| CorrT | 0.024 | 0.046 | 0.040 | 0.210 | 0.683 | 0.943 |
| CorrTRB | 0.025 | 0.044 | 0.048 | 0.228 | 0.710 | 0.943 |
| CorrTPB | 0.025 | 0.051 | 0.053 | 0.250 | 0.703 | 0.943 |
| CorrTMBG | 0.021 | 0.038 | 0.055 | 0.225 | 0.695 | 0.938 |
| CorrTMBR | 0.021 | 0.040 | 0.050 | 0.213 | 0.700 | 0.938 |
| CorrTMBM | 0.023 | 0.045 | 0.048 | 0.230 | 0.678 | 0.933 |

| | | | Heteroscedasticity | | | |
| | CvM | KS | $h = 0$ | $h = 5$ | $h = 10$ | $h = 15$ |
|---|---|---|---|---|---|---|
| CorrT | 0.013 | 0.037 | 0.053 | 0.240 | 0.708 | 0.960 |
| CorrTRB | 0.013 | 0.030 | 0.053 | 0.258 | 0.715 | 0.960 |
| CorrTPB | 0.013 | 0.026 | 0.065 | 0.270 | 0.705 | 0.965 |
| CorrTMBG | 0.016 | 0.035 | 0.053 | 0.265 | 0.705 | 0.958 |
| CorrTMBR | 0.018 | 0.039 | 0.058 | 0.258 | 0.685 | 0.955 |
| CorrTMBM | 0.012 | 0.026 | 0.065 | 0.255 | 0.688 | 0.955 |

The simulation results are collected in Table 5. We observe that all the methods perform quite well in that they all control the Type I error rate and thus inverting these tests would be a valid way of constructing confidence sets for the parameter under testing. This suggests that the bootstrap methods proposed by the authors can be extended to handle non-sparse high-dimensional linear models via the construction by [4]. In addition to the novel result by the authors who show that bootstrap methods can be successfully applied to high-dimensional sparse linear models, our simulation evidence suggests the possibility that the boundary of bootstrap methods might be further pushed to non-sparse settings although rigorous theoretical justification is left for future research. Another interesting direction would be investigating how to address the problem for $|G| > 1$ or even very large $|G|$ in non-sparse scenarios.

To observe finite sample differences between all of the methods we calculated Cramer-von-Mises (CvM) and Kolmogorov-Smirnov (KS) distances between the empirical null distribution of p-values and the uniform distribution. We see from Table 5 that CorrTMBG has the smallest distance from the uniform distribution in the case of homoscedastic errors. However, with regard to the heteroscedastic errors, all methods are practically indistinguishable with perhaps CorrTPB and CorrTMBM bootstrap performing slightly better. This last case mimics the low-dimensional property of the MBM and its strong robustness to the heteroscedasticity (see [2]).

Department of Mathematics and Rady School of Management, University of California, San Diego, La Jolla, CA, 92093